\documentclass[aps,prb,10pt,twocolumn,floatfix,superscriptaddress,showpacs,numerical,footinbib]{revtex4-1}

\usepackage{graphicx}
\usepackage{amsmath, amssymb, braket}
\usepackage[T1]{fontenc}
\usepackage[utf8]{inputenc}
\usepackage{bbold}
\usepackage[colorlinks=true,linkcolor=blue,citecolor=blue,breaklinks]{hyperref}
\usepackage[pdftex]{color}

\newcommand{\phd}{{\protect\vphantom{\dag}}}		% phantom dagger

\definecolor{purple}{rgb}{0.5,0,0.5}

\begin{document}
\title{Density matrix renormalization group on a cylinder \mbox{in mixed real and momentum space}}
\author{Johannes Motruk}
\affiliation{\mbox{Max-Planck-Institut f\"ur Physik komplexer Systeme, N\"othnitzer Str.\ 38, 01187 Dresden, Germany}}

\author{Michael P. Zaletel}
\affiliation{\mbox{Station Q, Microsoft Research, Santa Barbara, California 93106, USA}}

\author{Roger S. K. Mong}
\affiliation{\mbox{Department of Physics and Astronomy, University of Pittsburgh, Pittsburgh, Pennsylvania 15260, USA}}

\author{Frank Pollmann}
\affiliation{\mbox{Max-Planck-Institut f\"ur Physik komplexer Systeme, N\"othnitzer Str.\ 38, 01187 Dresden, Germany}}
\date{\today}
\begin{abstract}
We develop a variant of the density matrix renormalization group (DMRG) algorithm for two-dimensional cylinders that uses a real space representation in the direction along the axis of the cylinder and a momentum space representation in the direction around the circumference. 
%This mixed representation allows us to use the momentum around the circumference as a conserved quantity which significantly reduces the computational costs during the optimization,
The mixed representation allows us to use the momentum around the cylinder as a conserved quantity in the DMRG algorithm.
Compared with the traditional purely real-space approach, we find a significant speedup in computation time
%, better scaling with increasing DMRG bond dimension, 
and a considerable reduction in memory usage.
%As a by-product, it produces the momentum-resolved entanglement spectrum without further computation. 
Applying the method to the interacting fermionic Hofstadter model, we demonstrate a reduction in computation time by over 20-fold, in addition to a sixfold memory reduction.
%In addition, we show that, counter to expectations, the matrix product operator representation of the Hamiltonian is more efficient in the mixed basis.
%
\end{abstract}
\maketitle

\section{Introduction}

While the density matrix renormalization group (DMRG)\cite{W92,S2005,S2011} method was originally conceived as an algorithm for one-dimensional systems, it has shown tremendous success in exploring two-dimensional systems in recent years.\cite{SW12}
The 2D DMRG method uses geometries such as a cylinder of finite circumference so that the quasi-2D problem can be mapped to a 1D one.\cite{LP94}
Despite the development of genuinely two-dimensional tensor network optimization algorithms,\cite{V08,VMC08} DMRG is still a standard method due to its reliability and stable convergence properties.
Especially in the very active field of topological phases, it has been successfully applied to identify quantum spin liquids,\cite{YHW2011,JWB12}  fractional quantum Hall phases,\cite{SY01,S08,FRND08,K10,ZSH11,ZMP13,ZMPR15}
and bosonic and fermionic fractional Chern insulating states.\cite{ZKB13,CV13,GMZ15}

Despite the improvements and successes of DMRG for (quasi-)two-dimensional systems, calculations on cylinders or strips of large width remain an extremely challenging task.
The main cause for this is rooted in the behavior of quantum entanglement, which is governed by the area law:~\cite{CP2001,VLRK2003,CH2004,ECP2010} the computational cost in efficient DMRG implementations grows exponentially in the circumference (but not length) of the cylinder for gapped systems.
Progress has only been possible due to numerous  improvements to the original algorithm.
The inclusion of Abelian and non-Abelian symmetries,\cite{McCG2002,SPV2010,SPV2011,SV2012,Weichselbaum20122972} the introduction of single-site optimization with density matrix perturbation\cite{White2005,HMSW2015} and the development of real-space parallelization\cite{SW2013} have increased convergence speed and decreased the requirement of computational resources.
An infinite version of the algorithm\cite{M08} has facilitated the investigation of translationally invariant systems.
While the barrier to larger circumferences is exponential, the accuracy also increases exponentially with circumference in a gapped system, so further optimizations of the DMRG are both highly desirable and worthwhile.

%In the ground state of a gapped system, the entanglement between two subsystems scales with the boundary seperating the two parts; hence, in one dimension, it is size-independent whereas in two dimensions, it grows linearly with the system width.
% In DMRG calculations, the ground state is found variationally in the class of matrix prodcut states, states that can be represented by matrices of a finite dimension. This dimension 
%
%This high entanglement in 2D systems implies a dramatic increase of the bond dimensions of matrix prodcut states needed to faithfully represent the ground state which roughly scales exponentially with the width of the system. A further optimization of the efficiency of DMRG is therefore highly desirable.

In this paper, we introduce a modification of the 2D DMRG algorithm for fermionic systems on cylinders in which we represent the state in momentum space in the direction around the cylinder.
This allows momentum to be used as a conserved quantity and greatly reduces computational costs.
We test the algorithm for the interacting Hofstadter model\cite{H76} and report a speedup and better scaling of computation time with the bond dimension and drastically reduced memory usage.
At moderate DMRG bond dimensions ($\chi = 3200$) computational time is reduced 20-fold, and the advantage increases further with $\chi$.
Furthermore, we show that the efficient construction of the Hamiltonian as a matrix product operator (MPO) in our method results in a surprising \emph{reduction} of the MPO bond dimension compared to the traditional real space approach for the model under consideration.

The paper is organized as follows.
After introducing the concept of the algorithm in Sec.~\ref{sec:algorithm}, we present the numerical results comparing the real and mixed space approach in Sec.~\ref{sec:num}.
In Sec.~\ref{sec:MPO}, we give some technical details of the algorithm by reviewing the construction of Hamiltonians as matrix product operators (MPOs).
We detail the structure of an interacting Hofstadter MPO in the traditional real space formulation and in the mixed real and momentum space.
Finally, we conclude with a discussion in Sec.~\ref{sec:conc}.

\begin{figure}[t]
\includegraphics[width=\columnwidth]{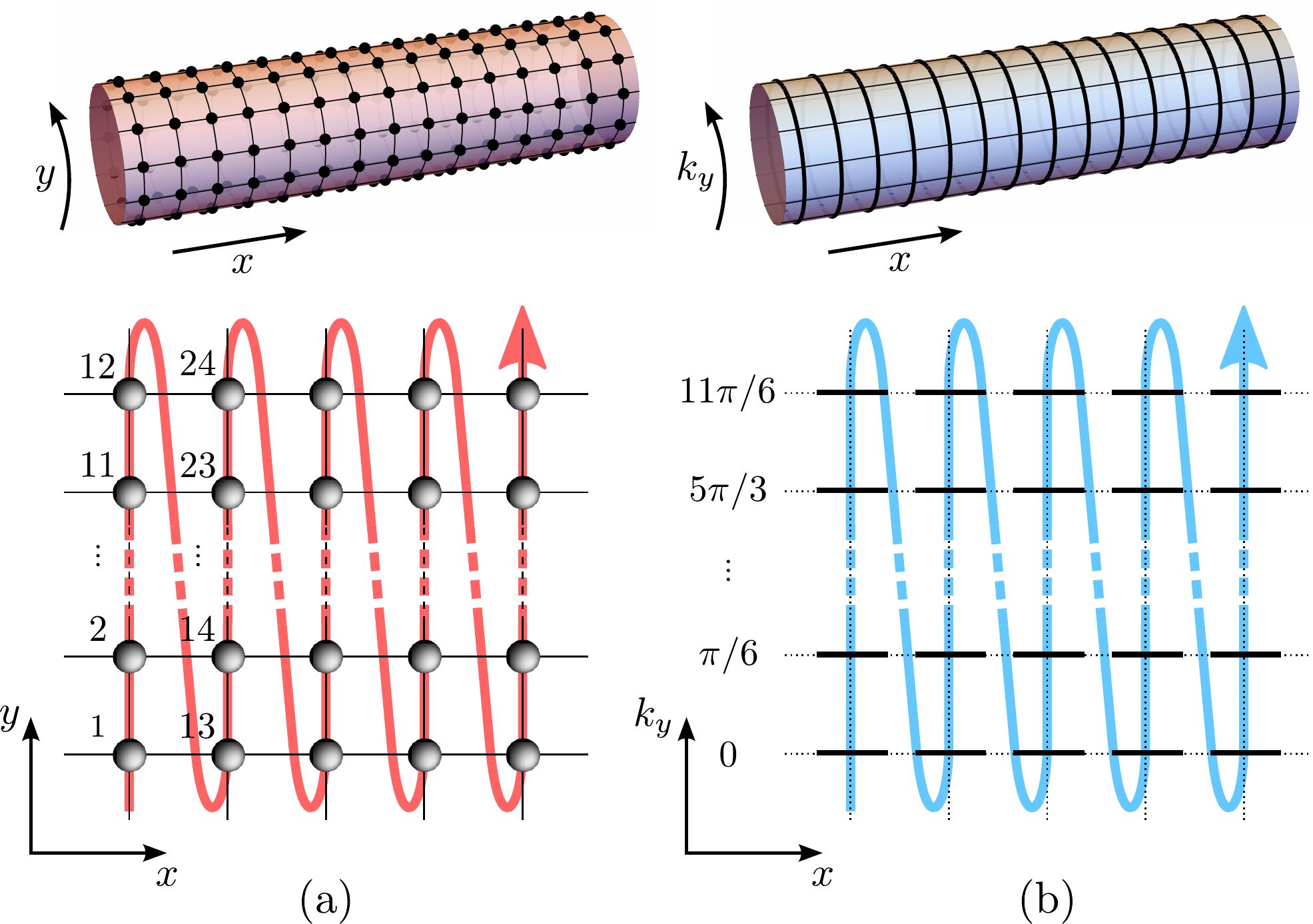}
\caption{Numbering scheme in order to transform the 2D lattice into a 1D chain. The horizontal direction is the direction along the cylinder, the vertical one around the cylinder. As an example, we show a cylinder of width $L_y=12$ (a) real space, (b) mixed real and momentum space approach. \label{fig:number}}
\end{figure}

\section{Concept of the algorithm \label{sec:algorithm}}

Cylinder DMRG requires mapping the Hilbert space of the  cylinder to a 1D chain with sites indexed by $i$. 
Generally this is done by letting the 1D chain ``snake'' through the real-space sites of the 2D cylinder, as illustrated in Fig.~\ref{fig:number}(a).
In order to simulate 2D cylinders while keeping computational costs manageable, it is crucial to exploit symmetries of the Hamiltonian.
To a exploit a global symmetry $\hat{R}_g$ in DMRG it must be onsite, meaning that its action factorizes across sites:
\begin{align}
\hat{R}_g = \prod_i \hat{R}^{(i)}_g. 
\end{align}
Charge conservation and spin-rotation are of this form.
However, when snaking in real space, a rotation of the cylinder \emph{permutes} the sites, so does not factorize as above, and cannot be used to accelerate computations.

To bring a rotation of the cylinder into an onsite form we use a mixed real-space and momentum-space basis.
In a fermion system, the Hilbert space of a real space site is spanned by the  occupations $n_{x, y} = 0, 1$ of the single-particle fermion orbitals.
We instead Fourier transform the single particle orbitals in the direction around the cylinder and pass to the many-body basis $\{ n_{x, k_y} = 0 , 1 \}$.
A crucial property of fermions is that the hard-core constraint is a direct consequence of the Pauli-exclusion, so remains true in momentum space.
This would \emph{not} be true in a bosonic system; projecting into the $n_{x, y} = 0, 1$ space does not lead to an equivalent restriction on $n_{x, k_y}$.
Thus we focus on fermion systems, though it would be interesting to develop a bosonic version for application to spin-systems.

The 1D chain  is again chosen to snake through the orbitals $i = (x, k_y)$ as shown in  Fig.~\ref{fig:number}(b).
While it is imperative that  $x$ remains ordered in the chain, to reduce the bipartite entanglement, there is freedom in choosing the ordering of the $k_y$. 
Indeed, different  orderings may require different intermediate DMRG bond dimensions (and hence different computational efficiencies) depending on the phase in question.\cite{LS03}
For example, if the state forms a charge density wave (CDW) at wavevector $Q_y = \pi$, it is advantageous to keep $k_y, k_y + \pi$ close together in the chain, as the CDW is dimerization in $k_y$ space.
In this work we study liquid phases, so we choose the simplest sequential ordering.

We note that while it is tempting to pass entirely to momentum space,\cite{Xiang96,Nishimoto2002} this would destroy locality in the $x$-direction.
The efficiency of the DMRG algorithm implicitly relies on short-range interactions along the length of the cylinder.
Even though eigenstates of non-interacting fermionic systems are product states in this basis, this behavior immediately changes once interaction terms are taken into account as the Hamiltonian becomes highly non-local in the momentum basis.\cite{ESLN15}

In the new ``mixed'' basis, a rotation of the cylinder takes the onsite form $T_y = \prod_{x, k_y} e^{i k_y \hat{n}_{x, k_y} }$, so can be exploited for a considerable speedup in computation time and a drastic reduction of memory usage by introducing an additional $\mathbb{Z}_{L_y}$ momentum quantum number where $L_y$ is the number of unit cells around the cylinder.
Furthermore, we readily obtain quantities such as the entanglement spectrum momentum-resolved without any further computation. We present numerical results benchmarking the mixed space algorithm versus the traditional real space approach for the interacting fermionic Hofstadter model in the following section.

\section{Numerical results \label{sec:num}}

\subsection{Hofstadter model}

To demonstrate the efficiency of our approach, we employ the infinite version of the algorithm (iDMRG) to calculate the ground states of the interacting Hofstadter model on an infinite cylinder geometry for different parameters.
The Hofstadter model describes fermions hopping on a square lattice subject to a magnetic field and the single-particle spectrum was shown to exhibit a fractal structure for arbitrary flux densities.\cite{H76}
In the case of rational flux densities  given by $\phi/\phi_0 = p/q$ per square plaquette, the spectrum separates into $q$ energy bands with non-trivial Chern number $C_i$.\cite{Zak64,TKNN82}
For non-interacting fermions, the model hosts incompressible states when an integer number of bands are fully occupied, and displays a Hall conductivity of 
\begin{equation}
 \sigma_{xy} = C \frac{e^2}{h},
\end{equation}
where $C = \sum_{i=1}^n C_i$ is the sum over the Chern numbers of the occupied bands.\cite{TKNN82}
More complex physics, however, arises in the case of a fractionally filled Chern band where lattice analogues of the fractional quantum Hall effect can emerge. These fractional Chern insulators (FCI) were theoretically proposed and numerically detected in numerous works for $C=1$\cite{MC09,NSCM11,SGS11,RB11} as well as for higher Chern numbers.\cite{LBF12,SRBR13,MC2015}

However, experimental realizations of FCI states remain elusive up to now. The Hofstadter model is of particular interest in this respect since its single-particle Hamiltonian has recently been realized in a system of ultracold atoms in an optical lattice\cite{Aid2013,MHK2013} and the Chern number of the lowest band has been experimentally determined to be unity.\cite{Aid2015}

\begin{figure}[t]
\includegraphics[width=\columnwidth]{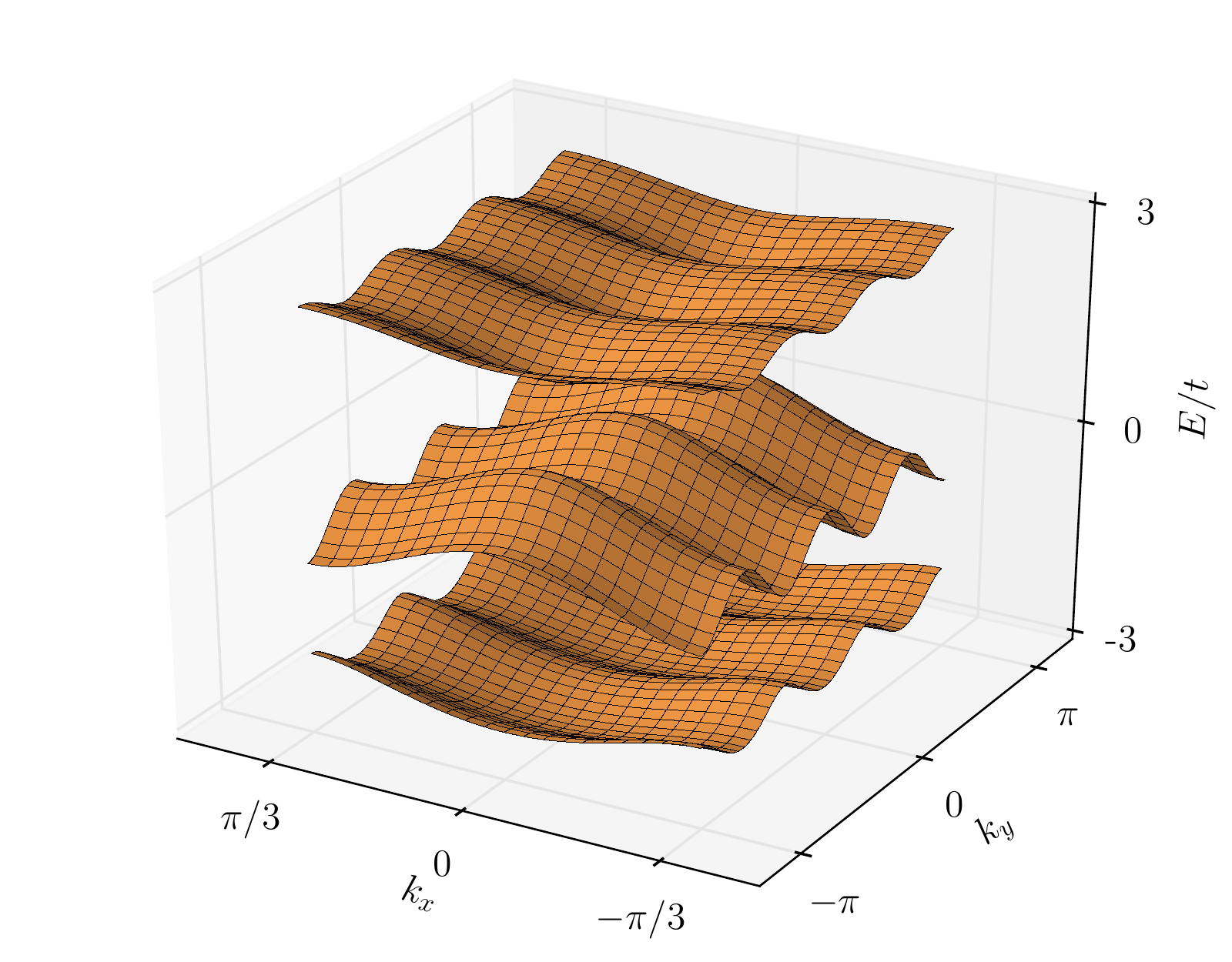}
\caption{Single-particle spectrum of the Hofstadter model for a flux density of $\phi/\phi_0 = 1/3$. The magnetic unit cell is chosen to be $l_x \times l_y = 3 \times 1$. \label{fig:bands}}
\end{figure}

\begin{figure}[b]
	\includegraphics[width=\columnwidth]{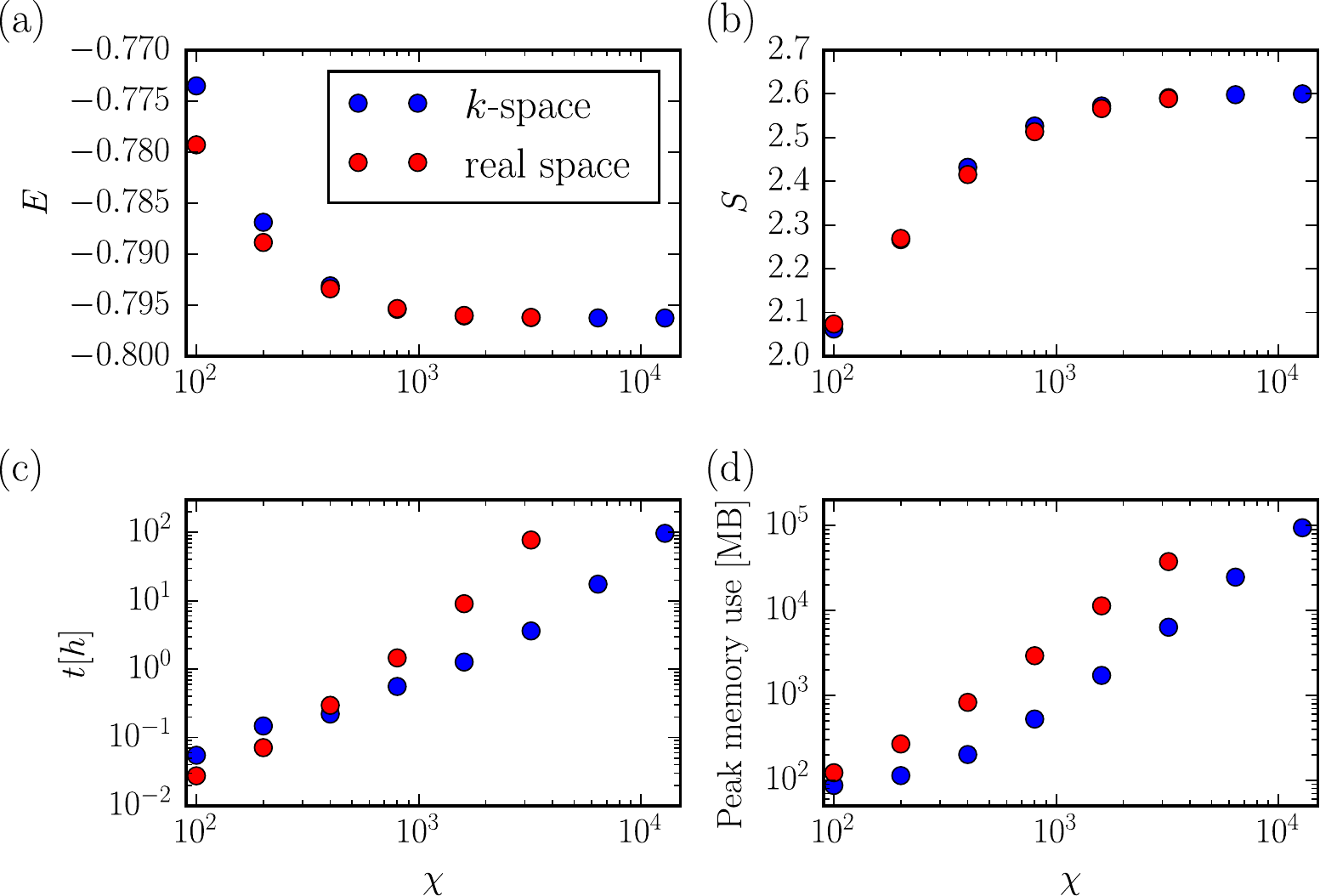}
	\caption{%
		Benchmark plots for the Hofstadter model on an infinite cylinder with $L_y=10$, $\phi = 2 \pi / 3$ and $V/t=0.1$ at one third filling.
		This corresponds to a fully occupied lowest band and weak interactions.
		Here $\chi$ is the DMRG bond dimension; increasing $\chi$ leads to better accuracy, at the expense of greater computational cost.
		The subplots are (a) energy, (b) entanglement entropy, (c) computing time, and (d) memory usage,
			as a function of $\chi$ for the range $\chi = 100 \mbox{--} 12800$.
	\label{fig:bench_full}}
\end{figure}

In the following, we will numerically investigate the Hofstadter model with Hamiltonian
\begin{equation}
	H =   -t \sum_{\langle j,l \rangle} \left( e^{i \phi_{jl}} c^{\dagger}_j c_l^\phd + \mathrm{H.c.} \right) + V \sum_{\langle j,l \rangle} n_j n_l \label{eq:H}  ,
\end{equation}
where $c_j^\dag$ ($c_j^\phd$) creates (annihilates) a fermion on site $j$, $\phi_{jl}$ is the phase acquired when hopping from site $l$ to site $j$ and $n_j$ is the particle number operator at site $j$.
By fixing a rational flux of $\phi = 2 \pi p/q$ per square plaquette and choosing the Landau gauge $\mathbf{A} = (0,Bx)$, the $\phi_{jk}$ are non-zero only for hoppings in $y$-direction and are invariant by a translation of $q$ sites in $x$-direction.
This allows for representing the single-particle problem in a magnetic unit cell (MUC) comprising $q \times 1$ sites and solving the $q$-site tight binding model via Bloch's theorem.
We then obtain $q$ energy bands with non-zero Chern numbers.
A typical plot of the band structure is given in Fig.~\ref{fig:bands}.
% First, we benchmark the $k$-space method by comparing the results to the ones obtained in real space.

\subsection{Fully occupied band with weak interactions}

To benchmark our algorithm, we first calculate the ground state of the model Hamiltonian \eqref{eq:H} for a fully occupied lowest band and weak interactions, i.e.~at one third site filling for the flux $\phi= 2 \pi /3$ per square plaquette and $V/t=0.1$. In this case, the lowest band of the non-interacting model has Chern number $C=1$ and hosts an integer quantum Hall state on the lattice when fully occupied which we expect to be stable against weak interactions.

The convergence of energy and entanglement entropy and the respective computing times and memory use with increasing DMRG bond dimension $\chi$ are depicted in Fig.~\ref{fig:bench_full}. In Fig.~\ref{fig:bench_full}(a) we observe that the energy of the $k$-space approach is higher for very low $\chi$ but this behavior quickly reverses. From $\chi=800$ on, the ground state energy obtained from the $k$-space method is lower than the one from the real space approach and in the converged region $\chi \geq 3200$, both methods yield the same energies.
We observe a similar behavior for the entanglement entropy $S$ between two semi-infinite halves of the cylinder in Fig.~\ref{fig:bench_full}(b). For $\chi<3200$, $S$ is slightly different between the two methods, but as expected, it converges to the same value with increasing bond dimension.

The reason for the different dependence of these quantities on the bond dimension for low $\chi$ originates from the different representation of the state. Since the direction around the cylinder is transformed into momentum space, the ``intra-ring entanglement'' within one ring of sites wrapping around the cylinder differs between the two methods. Thus even though both methods have the same ``inter-ring entanglement'' between two halves of the infinite system for a cut between rings, the difference in ``intra-ring entanglement'' has a small effect both for the entanglement entropy and the energy of the states at lower bond dimensions.

The behavior of these two quantities for low bond dimensions shows that the basis chosen for DMRG calculations can have an effect if $\chi$ is not high enough to fully represent the state of the system. This is particularly interesting at or near a critical point, where the state can never be faithfully represented by a matrix product state of finite bond dimension, though finite entanglement scaling\cite{Tagliacozzo-2008, FES2009} can provide information about the nature of the state.

In Fig.~\ref{fig:bench_full}(c), we show the computing time as a function of the bond dimension which clearly demonstrates the superiority of the $k$-space method compared to the traditional approach.
For higher bond dimensions, the computing time is significantly lower and it increases considerably slower as a function of $\chi$.
The presence of an extra conserved quantity provides an additional block structure to the tensors, severely reducing the computational cost, with  20-fold speed up at $\chi=3200$.
For the same reason the amount of memory needed to perform the calculations is dramatically reduced as depicted in Fig.~\ref{fig:bench_full}(d).
For bond dimensions $\chi>400$, the peak memory use in the mixed space approach is approximately six times lower.

\subsection{Partially filled band}

\begin{figure}[t]
	\includegraphics[width=8cm]{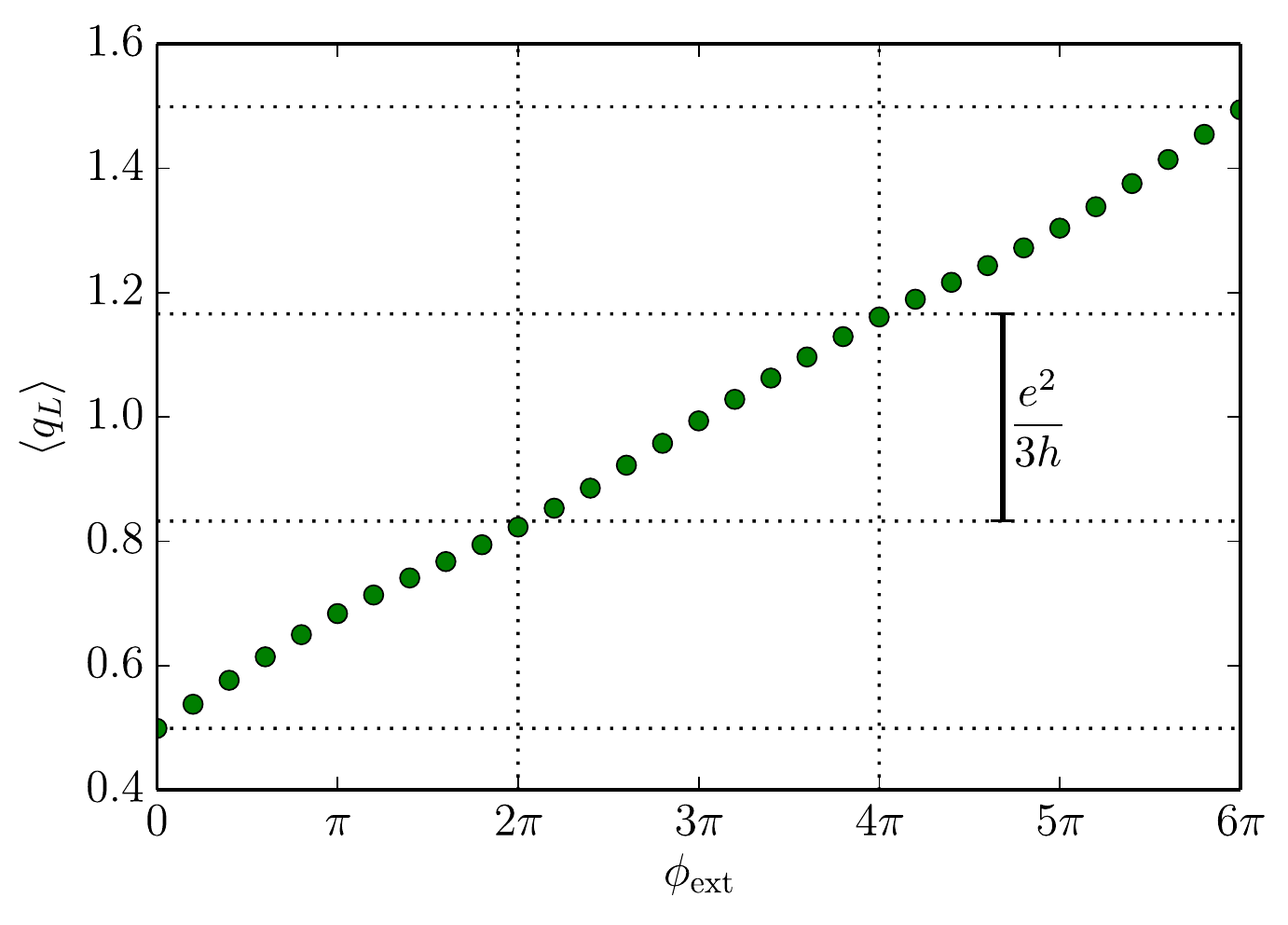}
	\caption{%
		Average charge value of the left Schmidt states as a function of flux going through the cylinder.
		After three flux quanta ($6 \pi$) have been adiabatically threaded through the cylinder, one charge has been pumped across the artificial cut of the system indicating a Hall conductivity of $\sigma_{xy}=1/3 \times e^2/h$.
	\label{fig:flux_pump}}
\end{figure}

Having tested the algorithm in the ``integer'' case in the previous paragraph, we now turn to the case of a partially filled band.
We compute the ground state of the Hamiltonian at one particle/nine sites, corresponding to one third filling of the lowest band at an interaction of $V/t=7.0$ with the mixed space algorithm.
Even though the interaction is significantly higher than the gap  $\Delta=2$ between the lowest and the second band, we expect the system to host  a fractional Chern insulator (FCI) state for these parameter values since FCI states have been shown to survive for interactions much larger than the band gap.\cite{GMZ15,KNC14} In order to confirm the presence of the FCI state, we compute the Hall conductivity $\sigma_{xy}$ of the system.
If there is a finite Hall conductance $\sigma_{xy}$, then adiabatically threading one flux quantum through the cylinder will pump the charge $\sigma_{xy}$ from the left to right half of the cylinder.\cite{L81} 
Thus, we adiabatically insert a flux through the system by twisting the boundary conditions and monitor the charge pumped into the left side of the system, as depicted in Fig.~\ref{fig:flux_pump}.
After a flux of $6 \pi$ has been inserted, a unit charge has been pumped across the cut showing a Hall conductivity of $\sigma_{xy}=1/3 \times e^2/h$.

\begin{figure}[t]
	\includegraphics[width=8cm]{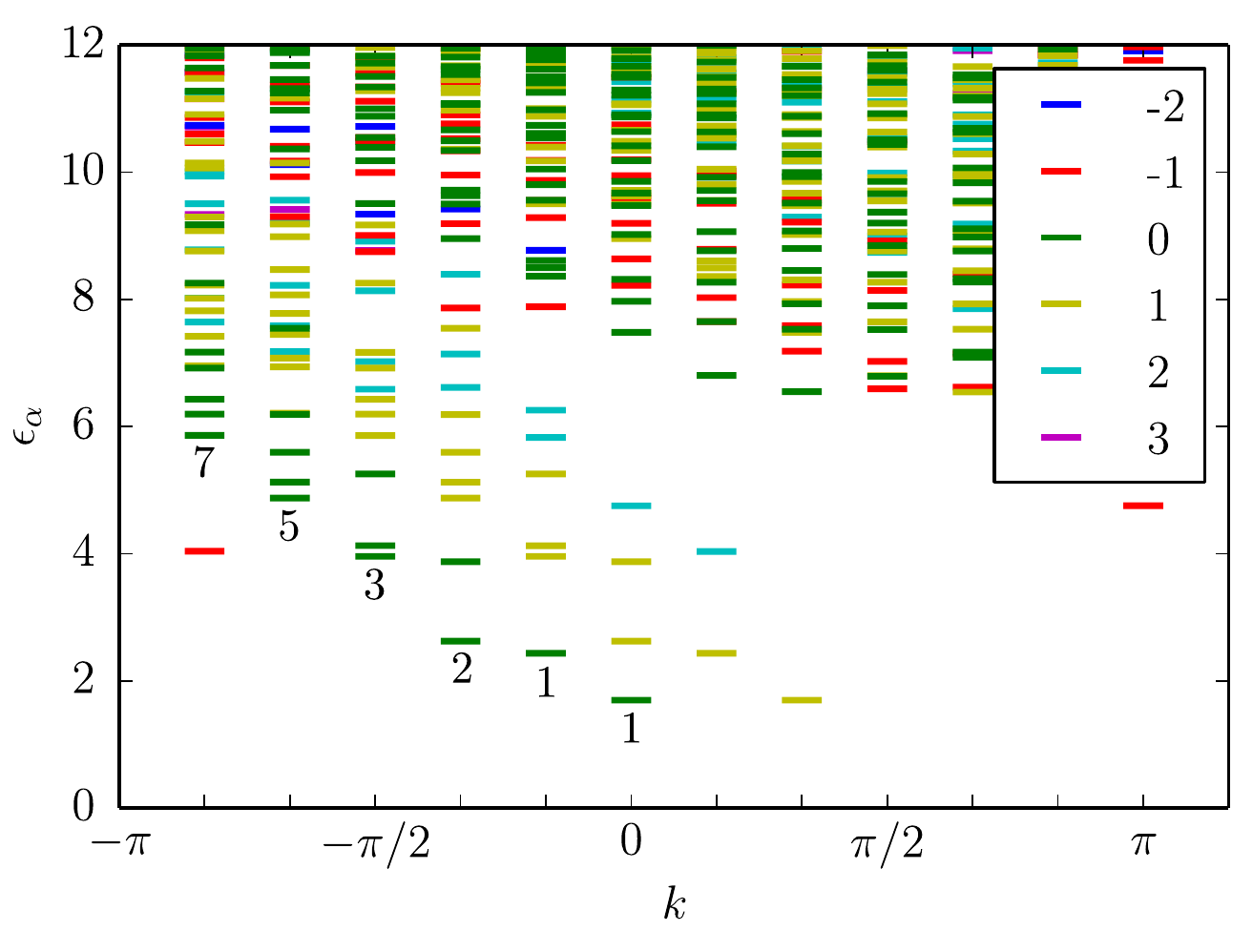}
	\caption{%
		Momentum resolved entanglement spectrum for $L_y=12$, $\phi = 2 \pi / 3$, $V/t=7.0$ and $\chi = 6400$ at one ninth filling (one third filling of lowest band). The different colors indicate the charge values of the corresponding entanglement eigenstates.
		The system is in a fractional Chern insulator phase and the chiral structure of the entanglement matching the edge theory state counting predicted from conformal field theory is clearly visible.
		The numbers indicate the counting for the zero charge sector depicted in green.
	\label{fig:ent_k}}
\end{figure}

Another advantage of the mixed space approach is the possibility of readily obtaining the momentum-resolved entanglement spectrum.
If we cut the system into two halves, we can represent the ground state $\ket{\psi}$ by a Schmidt decomposition given by
\begin{equation}
 \ket{\psi} = \sum_{\alpha} \lambda_{\alpha} \ket{\phi_{\alpha}}_L \ket{\phi_{\alpha}}_R
\end{equation}
where $\ket{\phi_{\alpha}}_{L/R}$ are the Schmidt states defined on the left and right side of the system.
These Schmidt states can be labeled by charge and $k_y$ momentum values.
The entanglement spectrum $\{ \epsilon_{\alpha} \}$ of the ground state can directly be read off the Schmidt decomposition via $\epsilon_{\alpha} = -2\ln \lambda_\alpha$.
%\noteRM{Does Fig.\ref{fig:ent_k} have the factor of two in $\epsilon_{\alpha} = -2 \ln \lambda_\alpha$?}\noteJM{no, it's without factor 2}
Whereas the calculation of the momentum labels of the Schmidt states requires additional computational steps in the real space basis,\cite{CV13} it is a trivial by-product of the algorithm in the mixed basis.
In this way, we readily obtain the entanglement energies labeled by momentum and charge values of the corresponding Schmidt states.

In the $\nu=1/3$ FCI state, we expect a structure in the entanglement energies matching the prediction of the corresponding conformal field theory for the edge.\cite{LH08,GMZ15} The counting pattern of the entanglement energy levels as a function of momentum for every charge sector should be $\{1,1,2,3,5,7,\ldots\}$.
This counting is clearly reproduced in Fig.~\ref{fig:ent_k} proving that the mixed real and momentum space algorithm correctly captures the $\nu=1/3$ FCI state in the Hofstadter model. 

Having demonstrated the suitability of our algorithm for evaluating the ground state of two-dimensional gapped systems on the example of the Hofstadter model, we turn to some technical details considering the representation of the Hamiltonian in the algorithm in the next section.

\section{Efficient MPO construction \label{sec:MPO}}

A crucial element of the (i)DMRG method is the representation of the Hamiltonian as a matrix product operator (MPO).\cite{VGC2004,Mc2007,S2011}
In this way, the operator acting on the infinite system can be expressed by a finite number of matrices that equals the number of sites in the iDMRG unit cell.
Here, we want to give a short pedagogical review about the construction of general MPOs and subsequently present the structure of the MPO of the mixed real and momentum space approach in some detail.

\subsection{Finite state machines}

Let us first consider a general operator acting on a chain of length $N$ as an introductory example.
Suppose the only terms are nearest neighbor couplings of the form
\begin{equation}
 \hat O = \sum_{i} \left( \hat A_i \hat B_{i+1} + \hat B_i \hat A_{i+1} \right) ,
 \label{eq:H_example}
\end{equation}
then $\hat O$ reads in tensor product representation as
\begin{align}\begin{split}
 \hat O &=
 	\hat A \otimes \hat B \otimes \mathbb{1} \otimes \cdots \otimes \mathbb{1}
	\\&\quad	+	\mathbb{1} \otimes \hat A \otimes \hat B \otimes \mathbb{1} \otimes \cdots \otimes \mathbb{1} + \cdots
	\\&\quad +	\hat B \otimes \hat A \otimes \mathbb{1} \otimes \cdots \otimes \mathbb{1}
	\\&\quad	+ \mathbb{1} \otimes \hat B \otimes \hat A \otimes \mathbb{1} \otimes \cdots \otimes \mathbb{1} + \cdots
	\label{eq:tensor}
\end{split}\end{align}
A pictorial way of writing down all summands of the operator is the representation of $\hat O$ in terms of a finite state machine (FSM).\cite{CB2008} 
A finite state machine consists of a set of states and a table of rules for transitions between the states.
An FSM can be depicted as a graph whose nodes (vertices) represent states and whose directed edges correspond to transitions between those states.
Conventionally, FSM are understood to be probabilistic, with the various possible transitions out of a state weighted probabilistically.
Each transition of the FSM into a new state has a corresponding action---for example appending a character to string--- so that by repeating sequentially a probability distribution over strings is built up. 
For our purposes these sequences will be taken in superposition, generating the summands of our Hamiltonian.
Therefore, the Hamiltonian is the sum of all possible transition paths generated by the FSM.
 
\begin{figure}[t]
	\includegraphics[width=8cm]{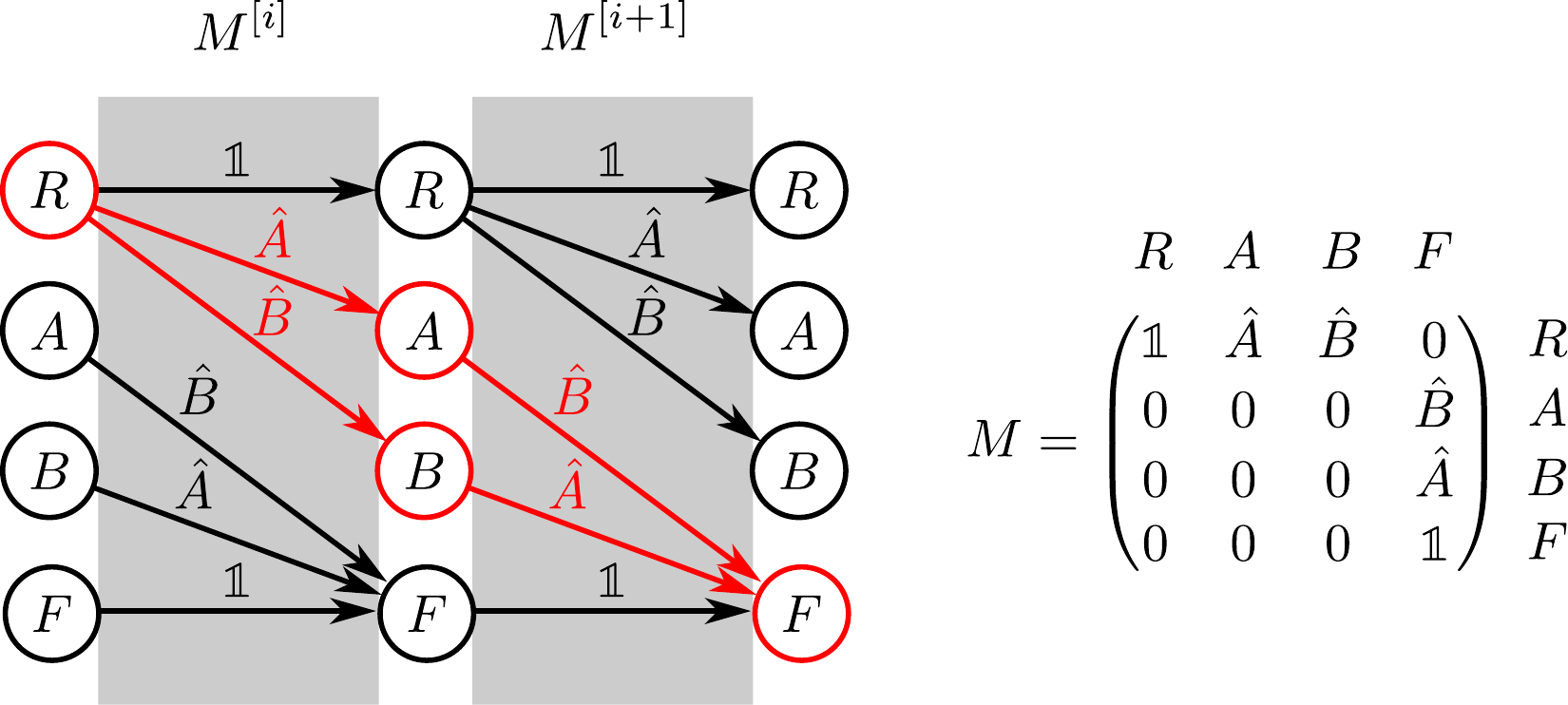}
	\caption{%
		Part of the FSM at sites $i$ and $i+1$ generating the operator $\hat O$ and MPO matrix for the matrix $M^{[n]}$ \eqref{eq:M_example}.
		The letters $R, A, B$ and $F$ label the states of the FSM as well as the rows/columns of the MPO matrix.
		Two paths of the FSM producing the term $\hat A_i \hat B_{i+1} + \hat B_i \hat A_{i+1}$ are highlighted in red.
		The gray rectangles indicate the six transitions in the FSM that exactly correspond to the six non-zero entries of $M$.
	\label{fig:fi_AB}}
\end{figure}

Here the transition on the $i$\textsuperscript{th} iteration of the FSM will place an operator on site $i$.
A part of the FSM generating the operator $\hat O$ is shown in the left illustration of Fig.~\ref{fig:fi_AB} and is to be read as follows.
We enter the FSM by starting in a ``ready'' state labeled by $R$.
From there, we follow all paths given by transitions between states leading to the ``final'' state labeled by $F$.
Each path represents one tensor product term in Eq.~\eqref{eq:tensor}.
When taking a transition between states, the operator which labels the transition is added to the tensor product. 

Let us now focus on a particular path generating the term $\hat A_i \hat B_{i+1}$.
It starts with a transition $R \rightarrow R$ in which the unit operator is added as the first term of the tensor product.
After going through $i-1$ of these transitions, the path jumps from $R$ into the state $A$ placing an operator $\hat A$ at the $i$\textsuperscript{th} site, and then from $A$ to $F$ adding $\hat B$ at site $i+1$ to the product.
From there on, it continues with transitions $F \rightarrow F$ adding unit operators until the tensor product has a length of $N$ operators.
%From this procedure, it is clear that all transitions correspond to sites in the chain, the $i$\textsuperscript{th} transition in the path to the $i$\textsuperscript{th} site of the chain.
The resulting operator is $\mathbb{1}_1 \otimes \cdots \otimes \mathbb{1}_{i-1} \otimes \hat{A}_i \otimes \hat{B}_{i+1} \otimes \mathbb{1}_{i+2} \otimes \cdots$ which is the desired term.

In Fig.~\ref{fig:fi_AB} all transitions corresponding to the sites $i$ and $i+1$ are depicted.
The entire operator $\hat O$ is created by taking a superposition of all paths in the FSM, corresponding to a sum of all tensor products.
It is easy to generalize the concept to an operator acting on an infinite chain in which the $R$ parts of the paths come from $-\infty$ and the $F$ parts of the paths go to $\infty$.
In this way, we may obtain a translationally invariant depiction of the FSM for any translationally invariant operator.

\subsection{MPO}
	
The representation of $\hat{O}$ as an FSM immediately leads to its representation as an MPO.
In the MPO formalism an operator $\hat O$ acting on the length-$N$ chain is written as
\begin{align}\begin{split}
	\hat{O} &= \sum_{a_0, \ldots, a_N} \vec v^\textrm{ left}_{a_0}
		M^{[1]}_{a_0 a_1} M^{[2]}_{a_1 a_2} \cdots M^{[N]}_{a_{N-1} a_N}
		\vec v^\textrm{ right}_{a_N} .
\end{split}\end{align}
Each $M^{[i]}_{a a'}$ is a physical \emph{operator} acting on site $i$, the indices $a,a'$ range from 1 to $D$, where $D$ is the number of states in the FSM picture.
Thus, it is convenient to interpret $M^{[i]}$ as a \emph{matrix of operators} on site $i$, in much the same way a matrix is used to represent a FSM or Markov chain.
The vectors $\vec v^\textrm{ left}$ and $\vec v^\textrm{ right}$ respectively initiate and terminate the MPO.

%\begin{align}\begin{split}
%	\hat O &= \sum_{\substack{j_1, \ldots , j_N \\ j'_1, \ldots , j'_N}} \vec v_{\rm left} M^{[1]j_1 j^\prime_1} M^{[2]j_2 j^\prime_2} \cdots M^{[N]j_N  j^\prime_N} \vec v_{\rm right}
%	\\	&\qquad	\times | j_1, \ldots , j_N \rangle \langle j^\prime_1, \ldots , j^\prime_N |,
%\end{split}\end{align}
%where $M^{[n]j_n j^\prime_n}$ are matrices of some finite dimension $D$ for finite range interactions and $\ket{j_n}$ and $\ket{j'_n}$ represent local states at site $n$.
%\noteRM{Why is there the qualifier ``finite range interactions''?}
%Here $D$ is the number of states in our FSM picture.
%In what follows, we write each $M^{[n]}$ as a $D\times D$ matrix of operators acting on site $n$.
%where each $M^{[n]}$ are matrices of some finite dimension $D$ for finite range interactions and $|j_n\rangle$ and $|j^\prime_n\rangle$ represent local states at site $n$.
%The vectors $\vec v^\textrm{ left}$ and $\vec v^\textrm{ right}$ respectively initiate and terminate the MPO.

By identifying the rows and colums of the MPO matrices with the states in the FSM as depicted in Fig.~\ref{fig:fi_AB}, we obtain the entries of the matrices $M^{[i]}$.
For example, the $i$\textsuperscript{th} transition from $R$ to $A$ places an operator $\hat{A}$ on site $i$ of the chain, and so $M^{[i]}_{R,A} = \hat{A}$.
This leads to the following matrices and intitiating/terminating vectors, written in the basis $(R,A,B,F)$:
\begin{align}
	M^{[n]} = \begin{pmatrix}
		\mathbb{1} & \hat A & \hat B & 0\\
		0 & 0 & 0 & \hat B \\
		0 & 0 & 0 & \hat A \\
		0 & 0 & 0 & \mathbb{1}
	\end{pmatrix}, \quad
	\begin{aligned}
		\vec v^\textrm{ left} &= (1,0,0,0),
	\\	\vec v^\textrm{ right} &= (0,0,0,1)^T.
	\end{aligned}
	\label{eq:M_example}
\end{align}
Multiplying these (taking tensor products of the operators), we obtain the sum of all terms of $\hat O$.
The concept can easily be extended to an infinite chain.
In the particular case of $\hat O$, the operator is invariant under the translation by one site and all the matrices $M^{[i]}$ along the chain are equal.
If an operator is only invariant under translation by $l$ sites, then there will be $l$ different MPO matrices along the chain.
In general, the dimensions of the matrices $M$ (which may vary on different sites), denoted $D$, are called the \emph{MPO bond dimensions}.

\subsection{Real-space MPO}

Let us now turn to the construction of the interacting Hofstadter Hamiltonian in MPO form.
%For the sake of easier notation, we will drop the index $y$ and just write $L$ for the circumference of the cylinder.
Working in the Landau gauge on an infinite cylinder, which guarantees translational invariance around the cylinder,
%We work in the Landau gauge and for clarity, we consider the case of a flux of $\pi$ per square plaquette since this realization has the smallest magnetic unit cell. However, the extension to different flux densities is straightforward.
the Hamiltonian in real-space is given by
\begin{equation}
 H = H_{\rm kin} + H_{\rm int} \label{eq:H_real} .
\end{equation}
The hopping and interaction terms are given by
\begin{align}\begin{aligned}
	H_{\rm kin} &= -t \sum_x \left( \sum_{y=1}^{L} c^{\dagger}_{x,y} c^\phd_{x+1,y} \right.
		\\&\;	+ e^{ i x \phi } c_{x,L}^\dag c_{x,1}^\phd
		\left. + \sum_{y=1}^{L-1} e^{ i x \phi } c_{x,y}^\dag  c_{x,y+1}^\phd \right) + \mathrm{H.c.} ,
	\\%\end{aligned} \\ \begin{aligned}
	H_{\rm int } &= V \sum_x \left( \sum_{y=1}^{L} n_{x,y} n_{x+1,y} \right.
		\\ &\qquad\qquad \left. + n_{x,L} n_{x,1} + \sum_{y=1}^{L-1} n_{x,y}  n_{x,y+1} \right) ~,
\end{aligned}\end{align}
where the index $x$ is the site labeling along the cylinder, $y$ labels the position around the cylinder and the flux per square plaquette is given by $\phi$.
For numerical implementation, we order the sites with increasing $x$, and then within each ring order by $y$ coordinates.
We then Jordan-Wigner transform the Hamiltonian according to
\begin{equation}
	c^\phd_i = \sigma^+_i \prod_{j<i} \sigma^z_j \quad \text{and} \quad c^{\dagger}_i = \sigma^-_i \prod_{j<i} \sigma^z_j.
	\label{eq:JW}
\end{equation}
This leads to strings of $\sigma^z$ operators between $\sigma^+$ and $\sigma^-$ in the hopping terms.

\begin{figure}[t]
	\includegraphics[width=\columnwidth]{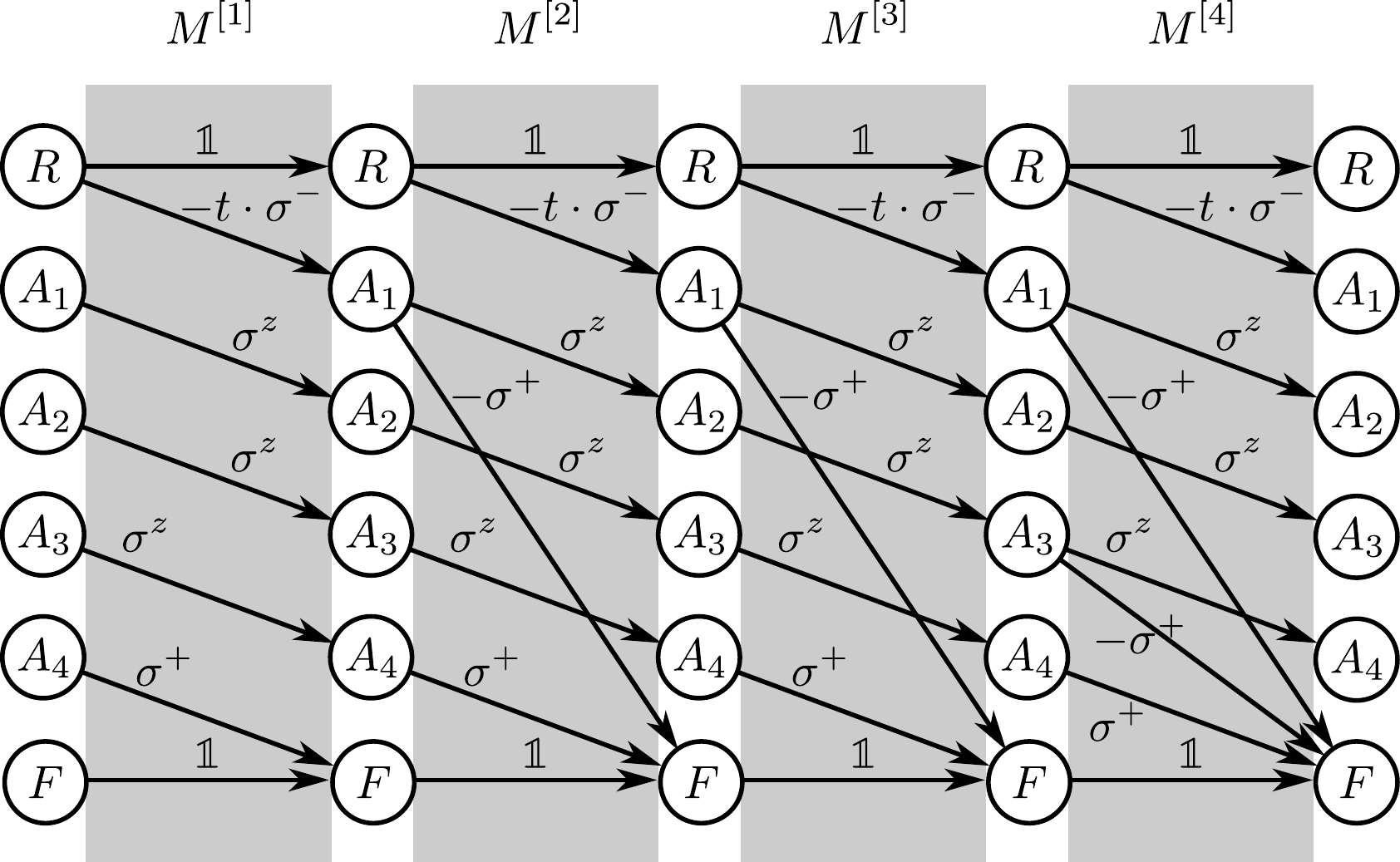}
	\caption{%
		Finite state machine creating the hoppings in the real space Hofstadter Hamiltonian for $L=4$ and $\phi = \pi$.
		For creating the Hermitian conjugate and interaction part of the Hamiltonian, we need two more copies of the depicted graph.
		In the Landau gauge, the Hamiltonian is invariant under translation by eight sites (two rings), which means that there are eight different MPO matrices.
		The remaining four matrices $M^{[5]-[8]}$ can be obtained from $M^{[1]-[4]}$ by simply replacing $-\sigma^+$ by $\sigma^+$ in the paths describing the hopping around the cylinder.
	\label{fig:finite_real}}
\end{figure}

As mentioned above, we have to construct $l$ MPO matrices to express the Hamiltonian in a unit cell of length $l$.
As an example, we show the finite-state machine creating the hopping terms of Hamiltonian \eqref{eq:H_real} for $L = 4$ and $\phi = \pi$, which has a unit cell of eight sites in Fig.~\ref{fig:finite_real}.
Note that this FSM will only create a part of the full Hamiltonian.
To obtain the Hermitian conjugate of the hoppings, an identical FSM with Hermitian conjugate operators is needed.
The interacting terms can be created by another copy of the graph without $\sigma^z$-strings and $\sigma^{+/-}$ replaced by $n$.
Furthermore, the four remaining matrices of the eight-site unit cell can be obtained by reversing the sign of the hoppings around the cylinder.
Putting together all finite state machines that produce the MPO matrices, the matrix dimension scales linearly with the circumference as $D=3L+2$.

\subsection{Mixed basis MPO}

\begin{figure}[t]
	\includegraphics[width=\columnwidth]{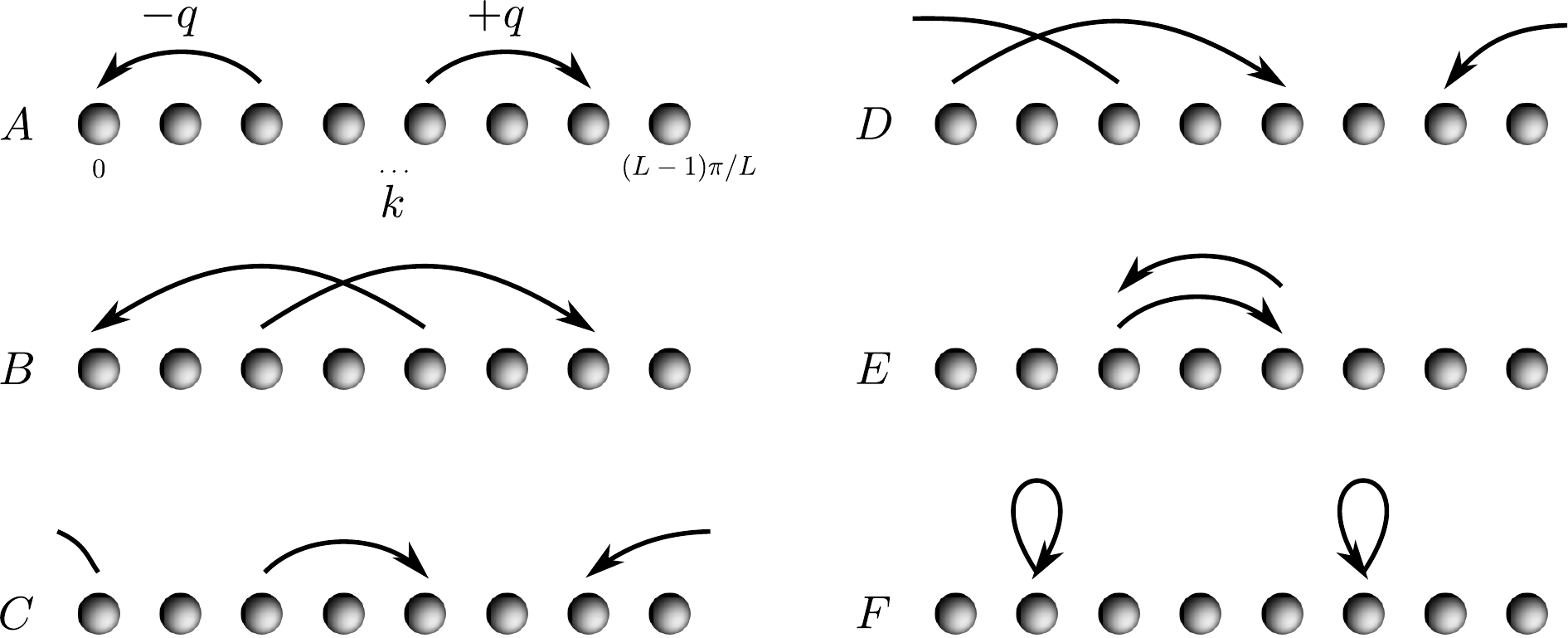}
	\caption{Different cases of momentum transfer within one ring occurring in the MPO due to the interaction term $H_{\rm int}$.
		Four cases not shown in the figure are related to $A$--$D$ via Hermitian conjugation.
	\label{fig:q_cases}}
\end{figure}

In order to use the momentum in the direction around the cylinder as a conserved quantity, we transform the Hamiltonian \eqref{eq:H_real} into $k$-space in the $y$-direction.
Then, the kinetic and interaction parts of the Hamiltonian read
\begin{align}\begin{split}
	H_{\rm kin} &= -t \sum_{x,k} \left[ \big( c_{x,k}^\dag c_{x+1,k}^\phd + \mathrm{H.c.} \big) \right.
	\\  &\qquad\qquad\qquad \left. + 2 \cos (k + x \phi) c_{x,k}^\dag c_{x,k}^\phd \right]
\end{split}\label{eq:H_kin}\end{align}
and
\begin{align}\begin{split}
	H_{\rm{int}} &= \frac{V}{L} \sum_{x,q} \Bigg[ (\cos q) n_{x,q} n_{x,-q}
		\;+\; n_{x+1,q} n_{x,-q} \Bigg]
%	\\ &\qquad\qquad + \sum_{k,k'} c_{x,k-q}^\dag c_{x,k}^\phd c_{x+1,k'+q}^\dag c_{x+1,k'}^\phd  \Bigg]
	\label{eq:H_k_int} ,
\end{split}\end{align}
with $k$ labeling the momentum in $y$-direction and $n_{x,q} = \sum_k c_{x,k+q}^\dag c_{x,k}^\phd$.
%The ``sites'' depicted around the cylinder now represent the different momentum eigenstates.
Since it only includes operators acting on one or two sites as in the real space version, the construction of the MPO part for the kinetic terms is straightforward.
The interaction part, however, consists mostly of terms with operators acting on four sites and additionally displays a momentum transfer dependent prefactor for interactions within one ring.

To demonstrate how the MPO is constructed, we focus on the interactions within the same ring
\begin{align}
	\frac{V}{L} \sum_{k,k',q} (\cos q) \, c_{x,k+q}^\dag \, c_{x,k}^\phd \, c_{x,k'-q}^\dag \, c_{x,k'}^\phd,
\end{align}
which represent the most complex terms in the MPO formulation. The remaining part of the MPO for all other terms may be constructed accordingly.
If we want to write this part in MPO form, every term has to be ordered according to the DMRG chain, with increasing momentum indices from left to right.
This leads to six distinct types of interaction terms within the ring, illustrated in Fig.~\ref{fig:q_cases}.
All other cases are either Hermitian conjugates or variants of $q \rightarrow L-q$ of these cases.

\begin{figure}[t]
 \includegraphics[width=\columnwidth]{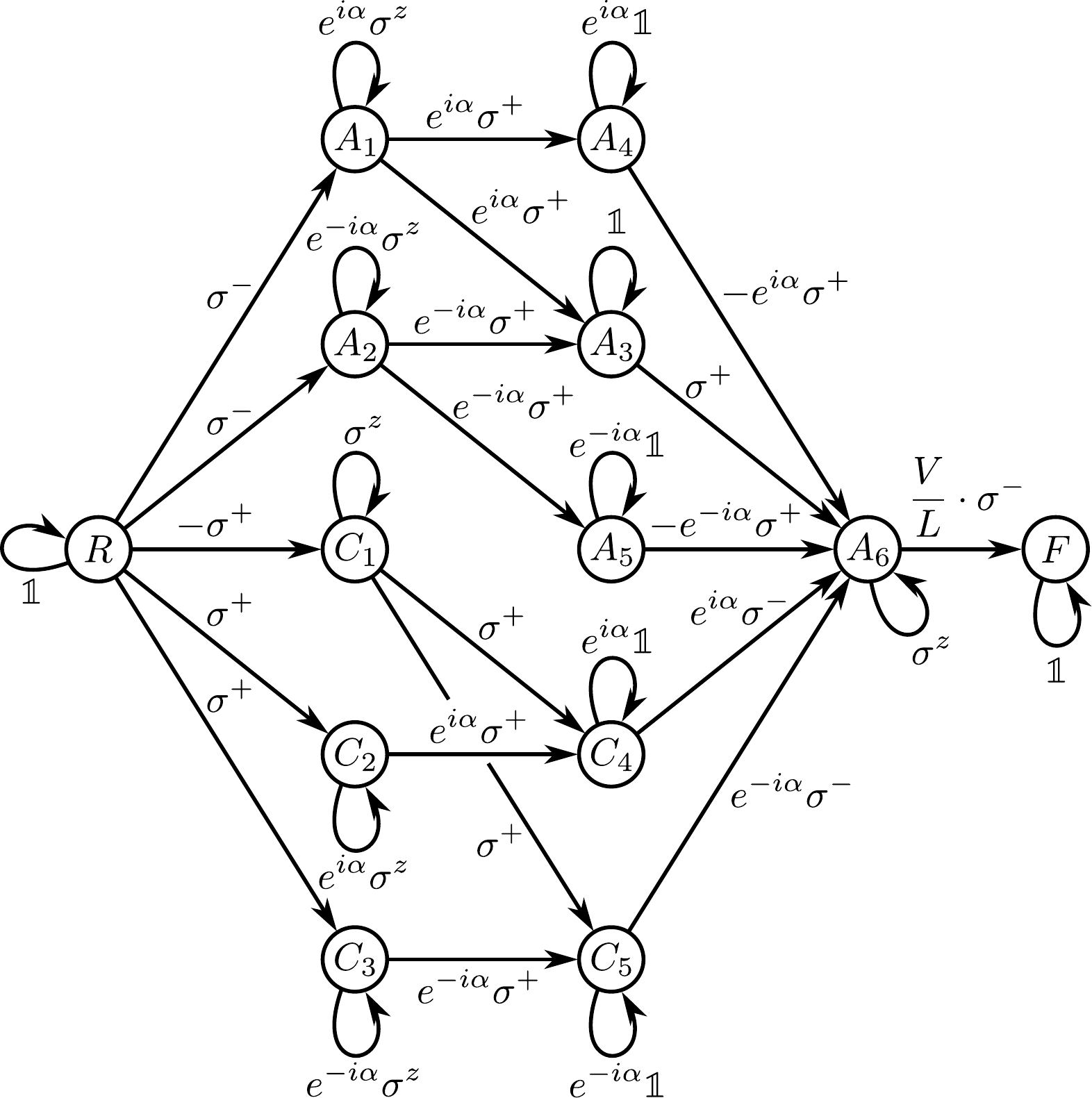}
	\caption{\emph{Coarse-grained} version of the FSM generating all intra-ring interaction terms of type $A$ to $D$ from Fig.~\ref{fig:q_cases}.
		The transitions between adjacent columns inserts a creation/annihilation operator at various momenta.
		The self-pointing ``loops'' places strings of $\mathbb{1}$/$\sigma^z$ operators in between the four creation/annihilation operators.
		In the actual FSM, there are multiple copies of the nodes $A_\mu$ and $C_\mu$ labeled by momenta in order to ensure momentum conservation (see Fig.~\ref{fig:k_path}).
	\label{fig:xk_mpo}}
\end{figure}

The terms of type $A$ and type $B$ take the form $c_k^\dag c_{k+l}^\phd c_{k+m}^\phd c_{k+m+l}^\dag$, with $m > l > 0$.
After the Jordan-Wigner transformation they become
\begin{align}
	\sigma_k^- \cdots \sigma_{k+l}^+ \times \sigma_{k+m}^+ \cdots \sigma_{k+m+l}^- ~,
\end{align}
	where the ellipsis denote a string of $\sigma^z$ for the intermediate sites.
These terms come with coefficients of $2[\cos l - \cos (m+l)]$ due to the $(\cos q)$ factor in the intra-ring interactions.
Type $C$ and $D$ terms are those that have two annihilation operators followed by two creation operators when ordered by momenta.
In the spin language, they take the form
\begin{align}
	\sigma_k^+ \cdots \sigma_{k+l}^+ \times \sigma_{k+l+m}^- \cdots \sigma_{k+l+m+n}^- ~,
\end{align}
with $l+2m+n = L$ to enforce momentum conservation.
The coefficients for these terms are $2[\cos(l+m) - \cos m]$.
Finally, type $E$ and $F$ terms are  number-number operator terms in $k$-space, of the form $2(1-\cos q) n_k n_{k+q}$.

\begin{figure*}[t]
	\includegraphics[width=\textwidth]{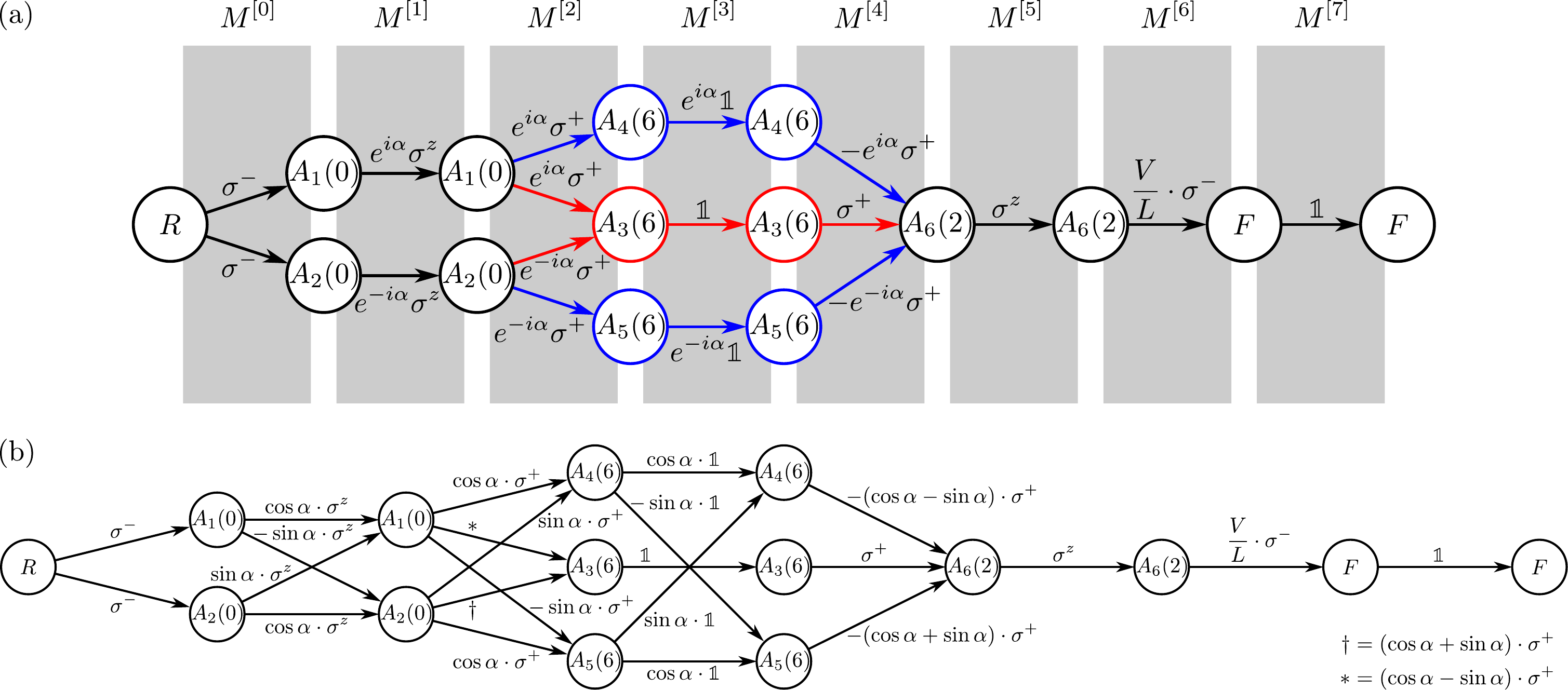}
	\caption{%
		(a) Application of the FSM from Fig.~\ref{fig:xk_mpo} for $L = 8$ and the specific term proportional to $c_0^\dag c_2^\phd c_4^\phd c_6^\dag$,
			which following the Jordan-Wigner transformation Eq.~\eqref{eq:JW} becomes $\sigma_0^- \sigma_1^z \sigma_2^+ \mathbb{1}_3^\phd \sigma_4^+ \sigma_5^z \sigma_6^- \mathbb{1}_7^\phd$.
		The red and blue portions of the path correspond respectively to case $A$ and case $B$ from Fig.~\ref{fig:q_cases}.
		Following the path from $R$ to $F$ generates the desired coefficient of $\frac{V}{L}2 (\cos 2\alpha - \cos 4\alpha)$ with $\alpha = 2\pi/L$.
		The intermediate nodes are labeled $A_\mu(k)$, where $k$ denotes the cumulative momentum, modulo $L$.  (Thus $6 \equiv -2 \pmod 8$) 
		(b) Modified path after rotating the basis which replaces the complex exponentials by real-valued sine and cosine terms.
	\label{fig:k_path}}
\end{figure*}

Figure~\ref{fig:xk_mpo} depicts a \emph{coarse-grained} version of the FSM which generates the type $A$--$D$ terms.
It is coarse-grained in the sense that if taken literally, it correctly captures only the distinct operator orderings which contribute to terms $A$--$D$, as well as the $q$-dependent prefactor $(\cos q)$  which is created by the phases $e^{\pm i \alpha}$ with $\alpha=2 \pi / L$.
However, it neglects the constraint placed by momentum conservation on the precise \emph{location} of the operator placements.
Implementing this constraint  will require duplicating the nodes $A_{\mu}, C_{\mu}$ to keep track of momentum conservation, as will be shown shortly.

We summarize how the various paths through the FSM produce the different types of terms from Fig.~\ref{fig:q_cases} in the following table: 
\begin{align*}\begin{array}{c|c @{\;\rightarrow\;} c @{\;\rightarrow\;} c @{\;\rightarrow\;} c @{\;\rightarrow\;} c}
	\textrm{Type} & \multicolumn{5}{c}{\textrm{path}}\\ \hline
	A  &  R  &  A_1/A_2  &  A_3      &  A_6  &  F \\
	B  &  R  &  A_1/A_2  &  A_4/A_5  &  A_6  &  F \\
	C  &  R  &  C_1      &  C_4/C_5  &  A_6  &  F \\
	D  &  R  &  C_2/C_3  &  C_4/C_5  &  A_6  &  F  
\end{array}\end{align*}
The FSM for type $E$ and type $F$ terms are not shown in the figure, but may be constructed with a similar idea.

% Type $A$ terms are generated via paths through the $A_3$ node, type $B$ terms via $A_4$ and $A_5$ nodes, type $C$ terms through $C_1$ node, and type $D$ terms via $C_2$ and $C_3$.
In the actual MPO implementation, there are multiple copies of each $A_\mu$ and $C_\mu$ nodes, one for each momentum quantum number (see Fig.~\ref{fig:k_path}).
Thus we label these copies by $k$,  $A_{\mu}(k), C_{\mu}(k)$.
The interpretation is that in state $X(k)$, the FSM has thus-far placed operators with total momentum $k$.
In this manner, the FSM can keep track of momentum conservation.
%One example of a collection of such paths is shown in Fig.~\ref{fig:k_path}.
%Going from the initial to final state in the FSM produces the interaction terms $A$ and $B$ depicted in Fig.~\ref{fig:q_cases}.
As a concrete example, Fig.~\ref{fig:k_path}(a) focuses on one specific subset of type-A paths in the FSM generating the terms proportional to $c_0^\dag c_2^\phd c_4^\phd c_6^\dag$, or equivalently $\sigma_0^- \sigma_1^z \sigma_2^+ \mathbb{1}_3^\phd \sigma_4^+ \sigma_5^z \sigma_6^- \mathbb{1}_7^\phd$ after a Jordan-Wigner transformation.
(The momenta are given in units of $2\pi/L$.)
The intermediate states $A_{\mu} (k)$ are labeled by the total momentum $k$ that has been placed along the path.
It is straightforward to verify that the graph yields the coefficient $\frac{V}{L}2 (\cos 2\alpha - \cos 4\alpha)$.

\begin{figure}[b]
	\includegraphics[width=\columnwidth]{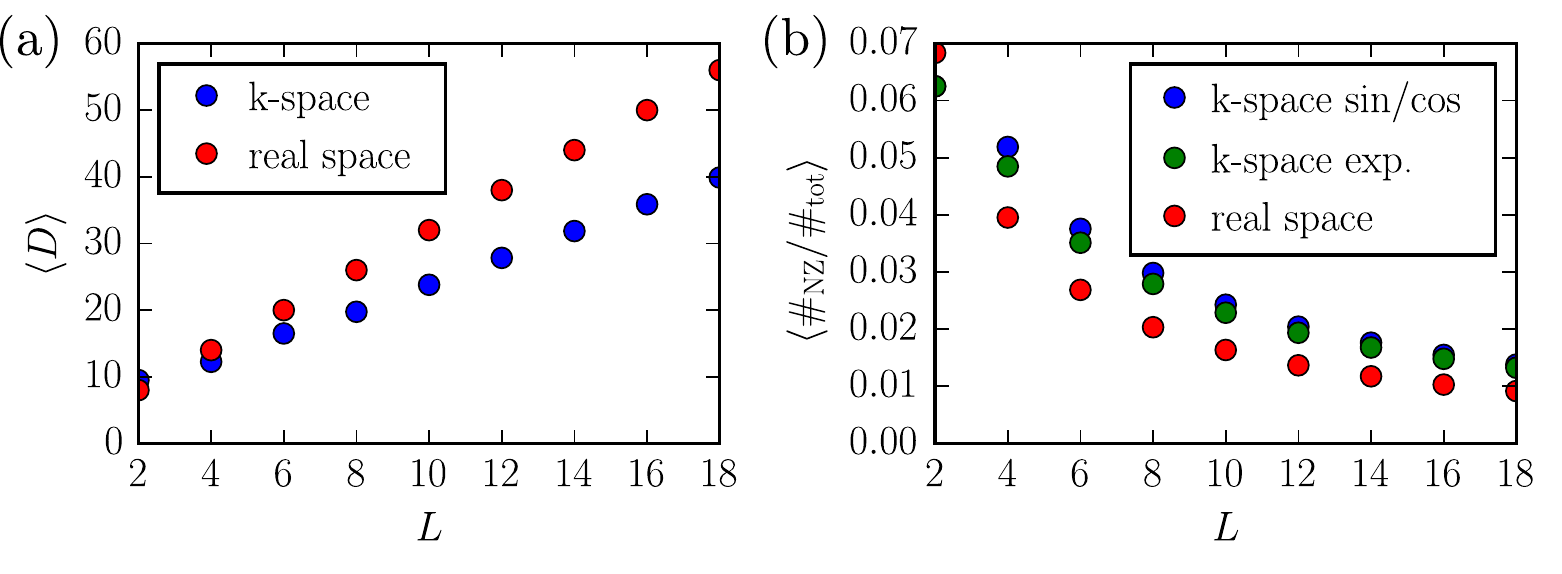}
	\caption{(a) Scaling of the average MPO bond dimension $\langle D \rangle$ with the circumference $L$ of the cylinder.
		Despite the additional complexity of the $k$-space MPO, its bond dimension is smaller than that of the real space MPO. (b) Average ratio of the number of nonzero vs.~total number of entries in the MPO matrices for the real space and the exponential and sine/cosine formulation in mixed space. 
	\label{fig:MPO_scale}}
\end{figure}

To account for all terms in the interacting Hamiltonian, naively we need more intermediate states than in the real-space formulation since first, the interaction part has a more complicated structure and second, the information about the momentum has to be encoded in the state.
If we assume each intermediate state type $A_\mu, C_\mu$ has an additional $L$-fold momentum label, a very crude upper bound for the MPO bond dimension is given by $D \leq 26L+9 $.
This is considerably larger than in the real space case, but still linear in $L$.
However, the true dimensions of the matrices in the computation are much lower.
Due to momentum conservation, many paths through the FSM are not allowed, and so at any specific site there are many unreachable intermediate states.
%In our implementation this `trimming' is all done automatically in a sub-routine for converting an FSM to an MPO, avoiding much tedium.
As a result, a significant fraction of such FSM states may be eliminated, resulting in much fewer states. In the MPO language this means that the MPO matrices comprise many rows or columns that do not contain any non-zero entries. We may simply delete these from the matrices. Note that in the real space version, no such rows or columns exist so that the size of the MPO cannot be reduced further.
In fact, after trimming the MPO dimension $D$ is significantly \emph{lower} in the mixed space than in real space. 
We show a scaling of the MPO bond dimension, averaged over the unit cell, as a function of the cylinder circumference in Fig.~\ref{fig:MPO_scale}(a).

As described above, the MPO includes complex numbers $e^{i \alpha}$.
However, $e^{i\alpha}$ and $e^{-i\alpha}$ always show up in pairs, as the matrix elements of $H_{\rm int}$ are real-valued.
Using the identity
\begin{align}
	\begin{pmatrix} e^{i \alpha}\! & 0 \\ 0 & \!\!e^{- i \alpha}\! \end{pmatrix}
	= \frac{1}{2} \!
	\begin{pmatrix} i & 1 \\ -i & 1 \end{pmatrix} \!
	\begin{pmatrix} \cos\alpha & \sin\alpha \\ -\sin\alpha & \cos\alpha \end{pmatrix} \!
	\begin{pmatrix} -i & i \\ 1 & 1 \end{pmatrix}\label{eq:rot}
\end{align}
we can further optimize the MPO.
By replacing the complex exponentials with the corresponding sine and cosine terms, we are able to perform our numerical calculations exclusively using real numbers. 
(Note that the hopping terms in Eq.~\eqref{eq:H_kin} are real-valued in the $k$-space basis.) The off-diagonal terms in the rotation matrix of Eq.~\eqref{eq:rot} generate more transitions in the resulting FSM, hence more nonzero entries in the MPO matrices, but do not increase the MPO bond dimension. To illustrate the effect of the transformation \eqref{eq:rot} on the FSM, we show the transformed version of the path from Fig.~\ref{fig:k_path}(a) in Fig.~\ref{fig:k_path}(b). Furthermore, we show the average sparsity of the different formulations in Fig.~\ref{fig:MPO_scale}(b) which may also serve as an indication of the computational complexity. In general, the MPO matrices in real space are slightly sparser than in the mixed representation, but the difference decreases with increasing circumference. The transformation in mixed space \eqref{eq:rot} does almost not affect the sparsity of the matrices.

\section{Conclusions \label{sec:conc}}
%\MZ{I think you can shorten this, no reason to review every detail (like technical implementation). Discussions and future direction more useful.}
% We have introduced a variant of the DMRG algorithm for fermionic lattice models on two-dimensional cylinders in which the direction along the cylinder axis is in real space  while the direction around the circumference has been transformed into momentum space.

The mixed space approach for DMRG on cylinders introduced in this paper uses the momentum around the cylinder as an additional conserved quantity, greatly reducing the numerical effort of the method.
Applying the algorithm to the interacting fermionic Hofstadter model, we have shown a speedup of up to 20 times in CPU time and up to 6 times reduced memory usage. In addition, the algorithm scales more favorably with the DMRG bond dimension $\chi$, offering the prospect of an even increased speedup for larger $\chi$.

%As a further advantage of the method, we obtain the momentum-resolved entanglement spectrum as a trivial by-product without having to perform further computational steps.

% Considering the technical implementation of the algorithm, we have reviewed the construction of matrix product operators (MPOs) by using finite state machines and shown that the MPO of the Hofstadter model possesses a more efficient representation in mixed space compared to the traditional real space formulation. The scaling of the MPO bond dimension with the cylinder circumference for general lattice models in mixed space is an interesting question to be investigated.

The drastically reduced computational cost suggests this approach could be a standard procedure when investigating fermionic lattice models with  DMRG. With quantum Monte Carlo suffering from the sign-problem in these systems, DMRG is still one of the most reliable algorithms to investigate ground states of fermionic systems when going beyond system sizes accessible in exact diagonalization.
An exciting future direction is to consider spinful Hubbard models in Mott insulating and Fermi-liquid regimes.\cite{noack}
Since the method allows us to reach much larger DMRG bond dimensions, a possible application is to detect exotic gapless non-Fermi liquid phases by finite entanglement scaling (FES).\cite{FES2009} The possibility of extracting the central charge of a critical phase from FES can reveal non-Fermi liquid behavior. \cite{Jiang2013,Geraedts2015}

The extension of the method to bosonic systems remains an open issue requiring further investigation. Since local interactions become non-local around the cylinder in the $k$-space representation, the onsite Hilbert space in mixed space has to be larger than two even for hard-core bosons. The increase in the dimension of the local Hilbert space will strongly affect the efficiency of the algorithm. For soft-core bosons where an artificial cutoff of the onsite Hilbert space has to be performed even in real-space, we also expect that the effective Hilbert space of a ``site'' in the mixed representation has to be larger to achieve the same accuracy of representing the state. While repulsive onsite interactions penalize a double occupancy in real space, this constraint is weakened due to the prefactor $1/L$ of the interaction terms in the mixed basis.

\acknowledgments
MZ and RM are grateful to the PKS visitor program.

\end{document}